\documentclass[final,3p,times]{elsarticle}
\usepackage{float,lscape}
\usepackage{graphicx}
\usepackage{parskip}
\usepackage{rotating}
\usepackage{amsmath}
\usepackage{longtable}
\usepackage{lscape}
\usepackage{caption}

\journal{Planetary and Space Science}

\biboptions{authoryear}

\begin{document}

\begin{frontmatter}

\title{Monte Carlo model for electron degradation in methane gas}

\author{ Anil Bhardwaj\corref{cor1}} 
\cortext[cor1]{Corresponding author: Phone: +91 471 2563663} 
\ead {anil\_bhardwaj@vssc.gov.in, bhardwaj\_spl@yahoo.com}
\author{Vrinda Mukundan\corref{cor2}}
\ead{vrinda\_mukundan@vssc.gov.in}
\address{Space Physics Laboratory,
Vikram Sarabhai Space Centre,
Trivandrum~695022, India}








\begin{abstract}
We present a Monte Carlo model for degradation of 1-10,000 eV electrons in an atmosphere of methane.
The electron impact cross sections for CH$_4$ are compiled and analytical  representations 
of these cross sections are used as input to the model.``Yield spectra'', which provides information  
about the number of inelastic events that have taken place in each energy bin, is used to calculate the yield (or population) of various 
inelastic processes. The numerical yield spectra, obtained from the Monte Carlo simulations, 
is represented analytically, thus generating the Analytical Yield Spectra (AYS).
AYS is employed to obtain the mean energy per ion pair and 
efficiencies of various inelastic processes. Mean energy per ion pair for neutral  
CH$_4$ is found to be 26 (27.8) eV at 10 (0.1) keV. Efficiency calculation
showed that ionization is the dominant process at energies $>$50 eV, for which 
more than 50\% of the incident electron energy is used. Above 25 eV, dissociation has an 
efficiency of $\sim$27\%. Below 10 eV, vibrational excitation dominates. Contribution of emission is
around 1.2\% at 10 keV. Efficiency of attachment process is $\sim$0.1\% at 8 eV 
and efficiency falls down to negligibly small values at energies greater than 
15 eV. The efficiencies can be used to calculate volume production rate in planetary atmospheres 
by folding with electron production rate and integrating over energy.
\end{abstract}

\begin{keyword}
Planetary atmospheres, molecular processes, Monte Carlo model, electron degradation, methane. 
\end{keyword}

\end{frontmatter}


\section{Introduction}

Electron collision with molecule can result in various processes, like 
ionization, dissociation, and excitation 
of the target molecule, which can produce new species that can be 
more reactive as well 
as physically and chemically different from their parent molecule. 
Secondary electron released during ionization can also initiate further 
reactions. By-products  of all these processes 
can initiate further reactions which are of great interest in the field of 
radiation chemistry, environmental chemistry,
planetary aeronomy processes, like aurora and dayglow, and also in astrophysical and 
biological systems 
\citep{Mason2003, Campbell12}. 
To understand such phenomena, a thorough knowledge of electron 
degradation when it collides with atoms or molecules is required.

Methane is the simplest hydrocarbon present in the solar system \citep{Wodarg2008}.
It causes infrared absorption in the atmosphere of Jupiter and Saturn 
and is an important atmospheric constituent in the
planets Uranus and Neptune \citep{Broadfoot1979}.
In Titan, photochemistry is governed by ionization and dissociation products of 
nitrogen and methane \citep{Lavvas2011}.
Collision of solar photons
or photoelectrons with methane molecules causes the neutral dissociation or 
ionization of the 
molecule which in turn leads to the generation of 
simple hydrocarbon radicals and ions. The subsequent reactions caused by these 
radicals and ions,
either with themselves or with methane and other background gases,  
cause the production of higher order hydrocarbons, be it alkanes, alkenes or 
alkynes \citep{Banaszkiewicz2000,Strobel2004} and leading to polymerization which
may produce UV-dark haze in auroral region of Jupiter \citep{Singhal1992} and very heavy ionic 
species in Titan's atmosphere \citep{Coates2007, Wahlund2009}. 
Hydrogen cyanide, an important precursor for the formation of amino acids and 
proteins, is formed from 
those reactions for which methane acts as a precursor \citep{Romanzin2005}. 
Fifth flyby of Titan by Cassini-Huygens mission found regions of low radar 
reflectivity which are interpreted as lakes, with methane as a major constituent \citep{Cordiar2009}.

The aim of this study is to present a Monte Carlo model which describes the 
degradation of electrons with energy in the range 1 eV to 10 keV in a CH$_{4}$ atmosphere. 
\citet{Gan1992} used solution of Boltzmann equations for studying degradation of electrons 
in CH$_4$ and they calculated energy transfer rates for elastic and various inelastic processes 
assuming a Maxwellian electron distribution.
Monte Carlo method is a stochastic method, which has 
been widely used for studying the problem of electron energy degradation in gases relevant for planetary atmospheres 
[\citet{Cicerone1971}, \citet{Ashihara78}, 
\citet{Green77}, \citet{Singhal80}, 
\citet{Singhal81}, \citet{Singhal91}, \citet{Bhardwaj93}, \citet{Bhardwaj99a}, 
\citet{Michael2000}, \citet{Bhardwaj99b}, \citet{Shematovich08}, 
\citet{Bhardwaj09}].
In this method, history of collisions of particles is simulated, and 
conclusions are drawn from the statistics of those histories.
Even though time consuming, at some levels it is found to be the most realistic 
simulations possible 
for studying electron energy deposition \citep{Solomon2001}. The energy loss process of electrons 
is actually discrete in nature and
this nature is exactly captured in the Monte Carlo model. The method make use of 
probabilistic decision making 
techniques, and accuracy of the result largely depends on the number of 
simulations carried out. 
The study involves two steps: compilation of cross sections for all  e-CH$_{4}$ 
collision processes 
and development of an  energy apportionment method to determine 
how electron energy is distributed in various loss channels.


\section{Cross Sections}
\subsection{Total and Differential Elastic} 

\begin{figure}
\noindent\includegraphics[width=0.60\linewidth]{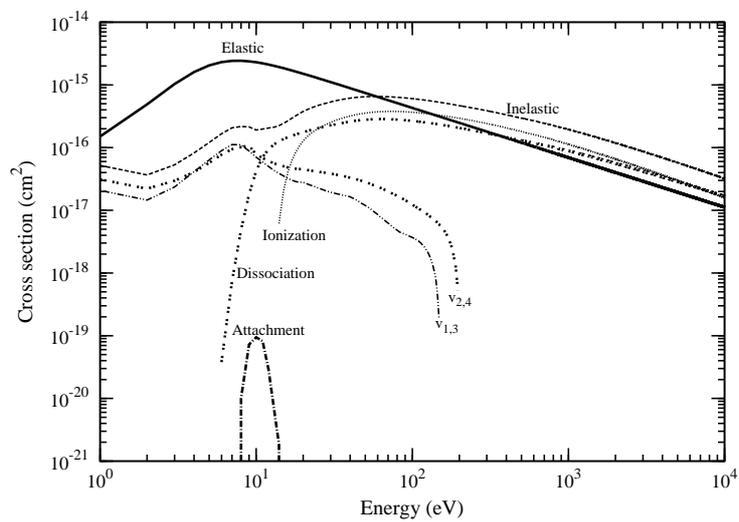}
\caption{Cross sections for elastic and inelastic processes for e-CH${_4}$ collisions.
${\nu}{_1,_3}$ and ${\nu}{_2,_4}$ are the cross sections for stretching and 
bending vibration modes, respectively.}
\label{total_xs}
\end{figure}

Total elastic scattering cross section for methane have been measured by \citet{Boesten91},
\citet{Bundschu97}, \citet{Iga99}, and \citet{Kanik93}. All these measurements 
are in good agreement with each other. Measurements of \citet{Boesten91} in the energy range 1.5-100 eV
was fitted using analytical formula by \citet{Shirai02}. At energies above 100 
eV, data of \citet{Kanik93} has been used for fitting. This analytically fitted form of 
elastic cross section is used in the current study and is shown in Figure \ref{total_xs}.

The direction in which the electron is scattered after collision with a 
CH$_{4}$ molecule is determined using differential elastic cross sections (DCS). DCS for e-CH${_4}$ 
collision has been measured by many workers. Values of DCS used in the present work are given 
in Table \ref{tab-EDCS}. 
In the low energy range of 3 to 15 eV, DCS measurements of \citet{Mapstone92} 
are used. However, DCS value at 5 and 10 eV are taken from \citet{Cho2008}. Cross sections 
for energies between 20 to 100 eV also  are  taken from \citet{Cho2008}. From 200 to 500 eV, measurements of 
\citet{Iga99} and at 700 eV  measurements of \citet{Sakae89} are used. Since DCS measurements are 
not available for CH$_{4}$ for energies greater than 700 eV, linearly extrapolated values 
of cross sections are used. 

\begin {landscape}
\begin{longtable}{p{0.4in}cccccccccl}
\captionsetup{singlelinecheck=off}
\caption{Elastic differential cross section for CH$_4$ in units 
of cm$^2$. Value inside the bracket indicates a 
         linearly extrapolated value. Notation 1E-18 implies 1 x 10$^{-18}$ }\\
	
        \hline
	Angles ($^{\circ}$) & 0 & 10 & 20 & 30 & 40 & 50 & 60 & 70 & 80 & \\
        Energy (eV) & & & & & & & &   \\
        \hline
        3.2 & (2.20E-17) &  (2.80E-17)  &     (3.40E-17)  &   
	4.00E-17   & 	4.60E-17   & 	5.40E-17   & 	9.40E-17   & 	8.10E-17 
	&  8.50E-17 &\\  
        
4.2 & (5.70E-17)  &  (6.40E-17)  &     (7.10E-17)  &   	7.80E-17   &	
8.50E-17   &	1.11E-16   &	1.19E-16   &	1.14E-16    &	1.01E-16 \\
5   & (8.23E-16)  &  (6.09E-16)  &      3.91E-16   &   	1.75E-16   &	
1.32E-16   &	1.28E-16   &	1.39E-16   &	1.48E-16    &	1.39E-16 \\
6   & (3.19E-16)  &  (2.73E-16)  &     (2.27E-16)  &   	1.81E-16   &	
1.35E-16   &	1.42E-16   &	1.30E-16   &	1.42E-16    &	1.38E-16 \\
7.9 & (7.32E-16)  &  (6.04E-16)  &     (4.76E-16)  &   	3.48E-16   &	
2.20E-16   &	1.80E-16   &	1.43E-16   &	1.39E-16    &	1.32E-16 \\
10  &  9.65E-16    & (8.03E-16)  &	6.40E-16   &   	4.88E-16   &	
3.51E-16   &	2.12E-16   &	1.77E-16   &	1.14E-16    &	0.86E-16 \\
15.4 & (7.65E-16)  & (6.29E-16)  &     (4.93E-16)  &   	3.57E-16   &	
2.21E-16   &	1.75E-16   &	1.08E-16   &	6.50E-17    &	5.00E-17\\
20   & (1.20E-15)  &  9.45E-16   &	6.91E-16   &   	4.55E-16   &	
2.69E-16   &	1.34E-16   &	0.83E-16   &	0.53E-16    &	0.36E-16\\
30   & (0.39E-15)  &  10.07E-16  &	6.25E-16   &   	3.36E-16   &	
1.60E-16   &	0.75E-16   &	0.40E-16   &	0.27E-16    &	0.18E-16\\
50   & (1.45E-15)  &  9.22E-16   &   	3.98E-16   &   	1.80E-16   &	
0.70E-16   &	0.29E-16   &	0.18E-16   &	0.10E-16    &	0.07E-16 \\
100  & (1.41E-15)  &  8.01E-16   &   	1.95E-16   &   	0.44E-16   &	
0.19E-16   &	0.09E-16   &	0.04E-16   &	0.03E-16    &	0.02E-16\\
200  & (1.14E-15)  &  5.51E-16   &	1.03E-16   &   	3.34E-17   &	
1.51E-17   &	8.20E-18   &	5.20E-18   &	3.80E-18    &	3.20E-18 \\
300  & (9.29E-16)  &  4.26E-16   &	7.26E-17   &   	2.56E-17   &	
1.12E-17   &	6.20E-18   &	3.80E-18   &	2.60E-18    &	1.80E-18 \\
400  & (6.76E-16)  &  3.02E-16   &	5.14E-17   &   	1.83E-17   &	
8.20E-18   &	4.00E-18   &	2.60E-18   &	1.50E-18    &	1.10E-18\\
500  & (6.46E-16)  &  2.80E-16   &	4.58E-17   &   	1.63E-17   &	
6.90E-18   &	3.20E-18   &	1.80E-18   &	1.20E-18    &	9.00E-19\\
700  & (1.54E-15)  &  2.19E-16   &	3.52E-17   &   	1.26E-17   &	
4.41E-18   &	1.98E-18   &	1.02E-18   &	6.47E-19    &	4.45E-19\\
800  & (2.38E-15)  & (1.94E-16)  &     (3.09E-17)  &   (1.12E-17)  &   
(3.53E-18)  &   (1.56E-18)  &   (7.69E-19)  &   (4.75E-19)   &  (3.13E-19)\\
900  & (3.67E-15)  & (1.71E-16)  &     (2.71E-17)  &   (9.74E-18)  &   
(2.82E-18)  &   (1.22E-18)  &   (5.78E-19)  &   (3.49E-19)   &  (2.20E-19)\\
1000 & (5.67E-15)  & (1.51E-16)  &     (2.37E-17)  &   (8.56E-18)  &   
(2.25E-18)  &   (9.64E-19)  &   (4.35E-19)  &   (2.56E-19)   &  (1.55E-19)\\
2000 & (2.56E-14)  & (6.66E-17)  &     (9.81E-18)  &   (3.66E-18)  &   
(5.09E-19)  &   (2.04E-19)  &   (6.70E-20)  &   (3.33E-20)   &  (1.54E-20)\\
3000 & (4.56E-14)  & (4.12E-17)  &     (5.85E-18)  &   (2.22E-18)  &   
(2.13E-19)  &   (8.27E-20)  &   (2.24E-20)  &   (1.01E-20)   &  (4.02E-21)\\
4000 & (6.56E-14)  & (2.93E-17)  &     (4.06E-18)  &   (1.56E-18)  &   
(1.15E-19)  &   (4.34E-20)  &   (1.03E-20)  &   (4.33E-21)   &  (1.54E-21)\\
5000 & (8.56E-14)  & (2.25E-17)  &     (3.05E-18)  &   (1.19E-18)  &   
(7.14E-20)  &   (2.64E-20)  &   (5.66E-21)  &   (2.25E-21)   &  (7.36E-22)\\
6000 & (1.05E-13)  & (1.82E-17)  &     (2.42E-18)  &   (9.52E-19)  &   
(4.83E-20)  &   (1.75E-20)  &   (3.46E-21)  &   (1.31E-21)   &  (4.01E-22)\\
7000 & (1.25E-13)  & (1.51E-17)  &     (1.99E-18)  &   (7.88E-19)  &   
(3.47E-20)  &   (1.24E-20)  &   (2.28E-21)  &   (8.36E-22)   &  (2.40E-22)\\
8000 & (1.45E-13)  & (1.29E-17)  &     (1.68E-18)  &   (6.69E-19)  &   
(2.61E-20)  &   (9.23E-21)  &   (1.59E-21)  &   (5.64E-22)   &  (1.54E-22)\\
9000 & (1.65E-13)  & (1.12E-17)  &     (1.44E-18)  &   (5.79E-19)  &   
(2.02E-20)  &   (7.09E-21)  &   (1.15E-21)  &   (3.99E-22)   &  (1.04E-22)\\
10000 &	1.85E-13  & (9.96E-18)  &     (1.26E-18)  &   (5.09E-19)  &   
(1.61E-20)  &   (5.60E-21)  &   (8.72E-22)  &   (2.93E-22)   &  (7.35E-23)\\	
\hline
\hline
\newpage
\hline
Angles (Degree) & 90 & 100 & 110 & 120 & 130 & 140 & 150 & 160 & 170 & 180\\
        \hline
	Energy (eV) & & & & & & & & & &  \\
        \hline
	3.2 & 8.30E-17  &    	6.20E-17   &   	3.40E-17    & 	2.20E-17 
        &2.00E-17   &    2.70E-17   &  	(3.40E-17)   &   (4.10E-17)   &  
(4.80E-17)   &  (5.50E-17)  \\  
4.2 & 1.01E-16  &	8.60E-17   &	7.30E-17    &	4.20E-17    &	2.50E-17   &	
5.30E-17   &	(8.10E-17)   &	(1.09E-16)   &	(1.37E-16)   &	(1.65E-16)  \\
5   &1.19E-16  &	0.84E-16   &	0.50E-16    &	0.24E-16    &	0.18E-16   &	
0.26E-16   &	0.36E-16   &	0.47E-16   &	0.60E-16   &	0.70E-16  \\
6   & 1.18E-16  &	8.10E-17   &	4.10E-17    &	2.20E-17    &	4.00E-17   &	
8.00E-17   &	(1.20E-16)   &	(1.60E-16)   &	(2.00E-16)   &	(2.40E-16)  \\
7.9 & 	9.80E-17  &	6.20E-17   &	2.70E-17    &	1.70E-17    &	4.60E-17   &	
9.40E-17   &	(1.42E-16)   &	(1.90E-16)   &	(2.38E-16)   &	(2.86E-16)  \\
10  & 0.55E-16  &	0.38E-16   &	0.24E-16    &	0.21E-16    &	0.31E-16   &	
0.58E-16   &	0.83E-16   &	1.20E-16   &	1.60E-16   &	1.78E-16  \\
15.4 & 4.00E-17  &	2.90E-17   &	2.70E-17    &	3.20E-17    &	4.30E-17   &	
5.70E-17   &	(7.10E-17)   &	(8.50E-17)   &	(9.90E-17)   &	(1.13E-16)  \\
20   & 0.26E-16  &	0.20E-16   &	0.19E-16    &	0.25E-16    &	0.32E-16   &	
0.40E-16   &	0.49E-16   &	0.56E-16   &	0.62E-16   &	0.69E-16  \\
30   & 0.12E-16  &	0.12E-16   &	0.13E-16    &	0.17E-16    &	0.21E-16   &	
0.27E-16   &	0.30E-16   &	0.32E-16   &	0.33E-16   &	0.34E-16  \\
50   & 	0.05E-16  &	0.04E-16   &	0.06E-16    &	0.08E-16    &	0.10E-16   &	
0.14E-16   &	0.16E-16   &	0.18E-16   &	0.19E-16   &	0.20E-16  \\
100   &	0.02E-16  &	0.02E-16   &	0.03E-16    &	0.03E-16    &	0.04E-16   &	
0.04E-16   &	0.05E-16   &	0.05E-16   &	0.05E-16   &	0.05E-16  \\
200  & 2.80E-18  &	2.40E-18   &	2.20E-18    &	2.00E-18    &  (2.15E-18)   &  
(2.25E-18)  &   (2.35E-18)  &   (2.45E-18)   &  (2.55E-18)   &  (2.65E-18)  \\
300  &	1.40E-18  &	1.20E-18   &	1.10E-18    &	1.10E-18    &  (1.00E-18)   &  
(1.00E-18)  &   (1.00E-18)  &   (1.00E-18)   &  (1.00E-18)   &  (1.00E-18)  \\
400 &	9.00E-19  &	8.00E-19   &	6.00E-19    &	6.00E-19    &  (6.00E-19)   &  
(6.00E-19)  &   (6.00E-19)  &   (6.00E-19)   &  (6.00E-19)   &  (6.00E-19)  \\
500 & 7.00E-19  &	6.00E-19   &	5.00E-19    &	5.00E-19    &	4.00E-19    &	
4.00E-19   &	4.00E-19   &	4.00E-19   &	4.00E-19   &	4.00E-19  \\
700   &	3.07E-19  &	2.46E-19   &	1.96E-19    &	1.68E-19    &	1.59E-19    &	
1.51E-19   &	1.42E-19   &	1.34E-19   &	1.25E-19   &	1.17E-19  \\
800 & (2.03E-19) &    (1.57E-19)  &   (1.23E-19)   &  (9.74E-20)   &  (1.01E-19)   &  
(9.28E-20)  &   (8.46E-20)  &   (7.75E-20)   &  (6.99E-20)   &  (6.33E-20)  \\
900   &  (1.34E-19) &    (1.01E-19)  &   (7.68E-20)   &  (5.64E-20)   &  (6.32E-20)   &  
(5.70E-20)  &   (5.04E-20)  &   (4.49E-20)   &  (3.91E-20)   &  (3.42E-20)  \\
1000  &  (8.92E-20) &    (6.46E-20)  &   (4.81E-20)   &  (3.27E-20)   &  (3.98E-20)   &  
(3.50E-20)  &   (3.00E-20)  &   (2.59E-20)   &  (2.18E-20)   &  (1.85E-20)  \\
2000 &  (6.13E-21) &    (3.41E-21)  &   (2.21E-21)   &  (9.06E-22)   &  (1.89E-21)   &  
(1.41E-21)  &   (9.88E-22)  &   (6.93E-22)   &  (4.66E-22)   &  (3.24E-22)  \\
3000 &   (1.28E-21) &    (6.11E-22)  &   (3.65E-22)   &  (1.11E-22)   &  (3.20E-22)   &  
(2.16E-22)  &   (1.34E-22)  &   (8.35E-23)   &  (4.93E-23)   &  (3.05E-23)  \\
4000 & (4.21E-22) &    (1.80E-22)  &   (1.01E-22)   &  (2.51E-23)   &  (9.06E-23)   &  
(5.71E-23)  &   (3.25E-23)  &   (1.85E-23)   &  (1.00E-23)   &  (5.70E-24)  \\
5000 &  (1.78E-22) &    (7.00E-23)  &   (3.78E-23)   &  (7.91E-24)   &  (3.40E-23)   &  
(2.03E-23)  &   (1.08E-23)  &   (5.79E-24)   &  (2.90E-24)   &  (1.55E-24)  \\
6000 & (8.80E-23) &    (3.23E-23)  &   (1.68E-23)   &  (3.08E-24)   &  (1.52E-23)   &  
(8.75E-24)  &   (4.42E-24)  &   (2.23E-24)   &  (1.05E-24)   &  (5.35E-25)  \\
7000 & (4.85E-23) &    (1.68E-23)  &   (8.49E-24)   &  (1.38E-24)   &  (7.77E-24)   &  
(4.28E-24)  &   (2.06E-24)  &   (1.00E-24)   &  (4.49E-25)   &  (2.18E-25)  \\
8000 & (2.89E-23) &    (9.54E-24)  &   (4.69E-24)   &  (6.95E-25)   &  (4.32E-24)   &  
(2.31E-24)  &   (1.07E-24)  &   (4.98E-25)   &  (2.14E-25)   &  (1.00E-25)  \\
9000 & (1.83E-23) &    (5.78E-24)  &   (2.78E-24)   &  (3.78E-25)   &  (2.58E-24)   &  
(1.33E-24)  &   (6.00E-25)  &   (2.69E-25)   &  (1.11E-25)   &  (5.03E-26)  \\
10000 &	 (1.22E-23) &    (3.70E-24)  &   (1.74E-24)   &  (2.19E-25)   &  (1.62E-24)   &  
(8.22E-25)  &   (3.57E-25)  &   (1.55E-25)   &  (6.21E-26)   &  (2.72E-26)  \\
\hline

       \label{tab-EDCS}
\end{longtable}

\end{landscape}

\subsection{Attachment}
Dissociative electron attachment process of  CH${_4}$ results in the 
production of H${^-}$ and CH${_2}{^-}$ ions.
Cross section for this process was measured by \citet{Sharp67} and \citet{Rawat07}. 
Former cross sections were analytically fitted by \citet{Shirai02}, which are used in the present work 
and is shown in Figure \ref{total_xs}.

\subsection{Vibrational Excitation}
Methane molecule is found to have four normal modes of vibration: ${\nu}{_1}$ 
with threshold energy 0.362 eV, ${\nu}{_2}$ with 0.190 eV,
${\nu}{_3}$ with 0.374 eV and ${\nu}{_4}$ with 0.162 eV. But it is difficult to 
resolve these modes experimentally as they have 
very close transition energies. Hence, experimental data are available for the 
combined cross section for symmetric ${\nu}{_1}$ and 
antisymmetric ${\nu}{_3}$ stretching vibrations (${\nu}{_1,_3}$), and symmetric 
${\nu}{_2}$ and 
antisymmetric ${\nu}{_4}$ bending vibrations (${\nu}{_2,_4}$) of CH${_4}$.

\citet{Shyn91} measured vibrational excitation cross sections for methane at 5.0, 7.5, 10.0 and 15 eV.
\citet{Tawara92} measured cross sections in the energy range 0.16-100 eV.  DCS values in the 0.6 to 5.4 eV 
range, measured by \citet{Bundschu97}, were integrated to obtain the integral cross section. 
Vibrational excitation cross sections used in the current study are taken from \citet{Davies88} in 
which cross section values are given for a larger energy range; 0.450 to 100 eV for ${\nu}{_1,_3}$ mode, 
and 0.162 to 150 eV for ${\nu}{_2,_4}$ mode. The values of \citet{Davies88} and \citet{Tawara92} agree well at 
energies 10-100 eV. However, at energies less than 10 eV there is a difference between two cross sections to a 
maximum of 50\% at few energies. Good agreement is 
found when cross sections of \citet{Davies88} are compared with that of \citet{Bundschu97}. Measurements of \citet{Shyn91} are 
found to be lower than the values of \citet{Davies88} by $\sim$50\%. Vibrational cross section used in our study 
is shown in Figure \ref{total_xs}.

\subsection{Ionization}

\begin{figure}
\noindent\includegraphics[width=0.60\linewidth]{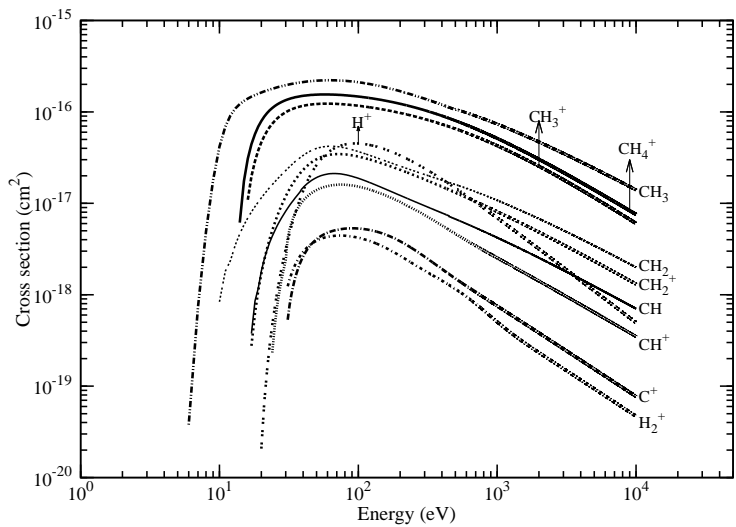}
\caption{Electron impact cross sections of CH$_4$ for various ionization and 
dissociation channels.}
\label{xs_channels}
\end{figure}

Ionization and dissociative ionization of CH${_4}$ results in the production of 
ions CH${_4}{^+}$, CH${_3}{^+}$, CH${_2}{^+}$ ,
CH${^+}$, C${^+}$, H${_2}{^+}$ and H${^+}$. Cross sections for these ionization 
processes have been measured by many authors, e.g., \citet{Tian97}, \cite{Chatham84}, 
\cite{Adamczyk66} and \citet{straub1997}. \citet{straub1997} have measured the cross
sections in the energy range 15-1000 eV and it is found to be the most reliable 
 among various available measurements \citep{Liu06}.
These measurements were later revised due to instrumental recalibration 
and was published in \citet{lindsay03} (Here after referred to as Straub's revised measurements).
\citet{Liu06} derived the oscillator strength and excitation functions
for various ionization channels of CH$_4$ and calculated the 
cross section values. These cross sections are in good agreement 
with Straub's revised measurements. However cross sections reported by \citet{Liu06} exclude
the contribution by pair production (e.g.(CH${_2}{^+}$, H${^+}$), (C${^+}$,H${^+}$)). \citet{erwin2008} using scaling law, developed analytical 
expressions for calculating cross sections for various ionization channels of methane, which are valid at 
all non-relativistic energies. These expressions allow calculation of the 
electron impact ionization cross sections  in an easier, and more direct way than
the functions derived by \citet{Liu06}. Good agreement is found between 
these theoretical ionization cross sections when compared with Straub's revised measurements and 
measurements of \citet{Tian97}. Maximum deviation ($\sim$20\%) is found for H$^+$ production.

For the present work, we have taken the cross sections 
for CH${_4}{^+}$ and CH${_3}{^+}$ production from \citet{Liu06}. For CH${_2}{^+}$ production, 
the cross section for positive ion pair formation (CH${_2}{^+}$, H${^+}$), measured
 by \citet{lindsay2001}, have been added with the values of \citet{Liu06} to account for the 
 contribution via doubly ionized channels. These cross sections are fitted using analytical equation \citep{Shirai02}; 
\begin{equation}
 \sigma = \sigma_oa_1(E/E_R)^{a_2}/[1+(E/a_3)^{a_2+a_4}+(E/a_5)^{a_2+a_6}] \label{ion-xs-eqn}
\end{equation}

where $\sigma_o$ = 1 x 10$^{-16}$ cm$^2$, E$_R$ is the Rydberg constant, and 
a$_1$, a$_2$,
a$_3$, a$_4$, a$_5$ and a$_6$ are the fitting parameters whose values are given 
in Table \ref{tab-ion-para}. The analytical expression of \citet{erwin2008} is used
 to calculate the cross sections for channels CH${^+}$, C${^+}$ and 
H${_2}{^+}$. For H${^+}$ production channel,
we have used the analytical expression of \citet{Shirai02}, extending it to
10 keV to get cross section values at higher energies. Figure \ref{xs_channels} 
shows the ionization and dissociative ionization cross sections used in the present study.

\begin{table}
\caption{Parameters for CH${_4}{^+}$, CH${_3}{^+}$ and CH${_2}{^+}$ ionization 
cross section (equation (\ref{ion-xs-eqn})) } 
\label{tab-ion-para}
\begin{tabular}{cccccccc}
\hline
Process & E$_{th}$(eV) & a$_1$ & a$_2$ & a$_3$ & a$_4$ & a$_5$ & a$_6$\\
\hline
CH${_4}{^+}$ & 12.99  & 4.40 & 1.627 & 7.720E-3 & -4.50E-2 & 
3.10E-2& 0.93 \\
CH${_3}{^+}$ & 14.24  & 2.18 & 1.435 & 1.13E-2& 7.4E-2& 
4.91E-2& 1.01 \\
CH${_2}{^+}$ & 15.20  & 0.121 & 1.868 & 3.44E-2& 3.00E-1 & 
5.20E-2& 0.91\\
\hline
\end{tabular}
\end{table}

\subsection{Dissociation}

Dissociation of methane by electron impact results in the production of 
neutral radicals CH${_3}$, CH${_2}$ and 
CH. Experimental cross sections for these processes are not available over a 
wide energy range, except for the CH${_3}$
radical production where the measurements are made up to 500 eV 
\citep{motlagh1998}. \citet{erwin2008} have given analytical expression for CH$_4$ 
dissociation cross sections which is valid at all non-relativistic energies. 
But the analytical expression does not 
account for the production of CH$_3$ radical through dissociative ionization channel 
CH${_3}$ + H$^+$. For CH${_3}$ radical production, we have used the analytical
 representation of \citet{motlagh1998} cross sections, as given by \citet{Shirai02} and 
 extended it to 10 keV. For CH${_2}$ and CH radical production, cross section are calculated using
the analytical expression of \citet{erwin2008}. Figure \ref{xs_channels} 
shows the dissociation cross sections used in the study.

\subsection{Emission}

\begin{figure}
\noindent\includegraphics[width=0.60\linewidth]{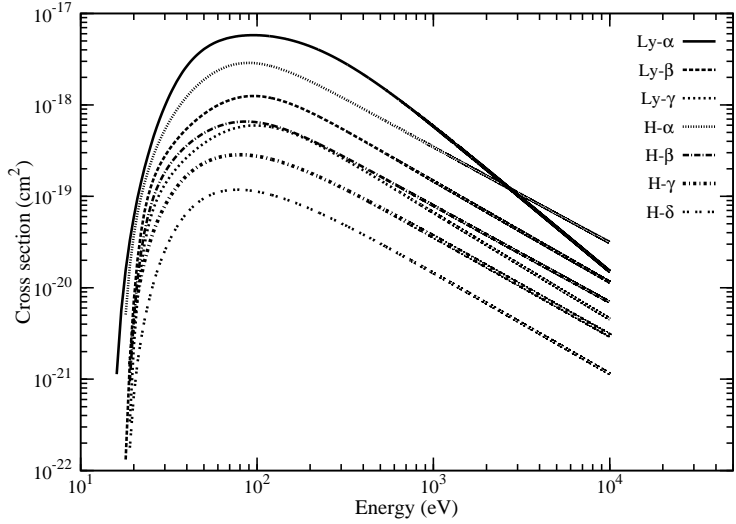}
\caption{Electron impact cross sections for H Lyman and H Balmer emissions.}
\label{xs_emission}
\end{figure}

Electronically excited state of neutral CH${_4}$ leads to the dissociation of 
the molecule resulting 
in the production of excited fragments \citep{Danko2011}. Cross sections for hydrogen 
 lyman series and carbon lines for energies less than 400 eV was measured by \citet{Pang1987}. \citet{Motohashi1996} 
measured cross sections 
for the emission from various excited fragments : the hydrogen Lyman and 
Balmer series, CH band emission at 420-440 nm,
line emissions from C at 165.7 nm and 156.1 nm and these cross sections are found to be in agreement with that of \citet{Pang1987}. The uncertainties in these cross sections were 
estimated to be $\pm$20\% for 
H Lyman-$\alpha$, $\pm$12\% for H Balmer-$\alpha$, $\pm$20\% for 
CH band, and $\pm$50\% 
for atomic carbon emission. Maximum energy of experimentally measured cross sections are 1 keV, 6 keV, 5 keV and 
1 keV for H Lyman-$\alpha$, H Balmer-$\alpha$, CH band, and C emissions, respectively. Analytic cross 
sections for these emission 
processes given by \citet{Shirai02} are 
extended upto 10 keV and used in the current model. (cf. Figure \ref{xs_emission})\\

\begin{figure}
\noindent\includegraphics[width=0.60\linewidth]{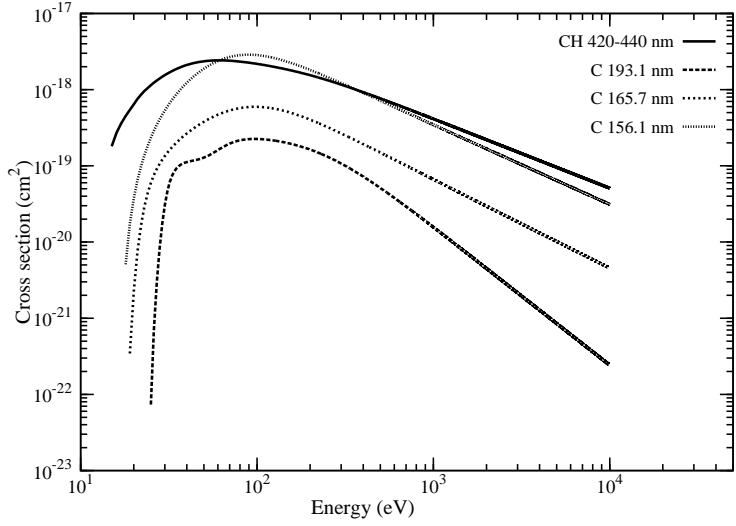}
\caption{Electron impact cross sections of CH$_4$ for CH band and various 
C I line emissions.}
\label{xs_emission2}
\end{figure}

The total inelastic cross section is obtained by adding the cross sections of above 
mentioned inelastic processes. 
Cross sections for the emission process and H$^+$ production channel are not taken into 
account while calculating total inelastic
cross section as they are already accounted in other channels. Figures \ref{total_xs}, 
\ref{xs_channels}, \ref{xs_emission} and \ref{xs_emission2}
shows the cross sections for elastic and various inelastic processes of methane 
that have been used in the present work.  

\section{Monte Carlo Model}

A model for local degradation of electrons in CH${_4}$ gas is developed in 
the energy range 1 eV to 10 keV using the Monte Carlo technique. Electrons incident upon CH${_4}$ molecules  
 deposit 
their energy into the gas through elastic or inelastic collisions. 
Inelastic collisions lead to further ionization, dissociation, excitation or 
attachment processes. Thus, there exists a range of possible channels through which an electron can 
degrade its energy on colliding with 
molecules. In the model, every incident electron is followed in a 
collision-by-collision manner until its energy goes below an assigned value. 

The Monte Carlo simulation starts after the  initial energy of electron is assigned. 
The direction of the incoming particle ($\theta$, $\phi$) is decided using 
random numbers R$_1$ and R$_2$ as 

\begin{equation}                             
\theta=\cos^{-1}(1-2R_1),           
\end{equation}
\begin{equation}                       
\phi=2\pi R_2.                     
\end{equation}
The distance to next collision is calculated from
\begin{equation}
S = -\log(1-R_3)/n\sigma_T,        
\end{equation}

where R$_3$ also is a random number and n is the number density of the
target particles, which is taken as 10$^{10}$ cm$^{-3}$.
 $\sigma_T$ is the total electron impact collision cross section, which is 
 \begin{equation}
 \sigma_T = \sigma_{el} + \sigma_{in} 
\end{equation}

where $\sigma_{el}$ and  $\sigma_{in}$ are the total elastic and inelastic 
collision cross sections. 
Decision on the type of collision that occur is made by comparing the 
probabilities of elastic and inelastic collisions,
P$_{el}$ and P$_{in}$, with another random number generated, R$_{4}$, where 
P$_{el}$ and P$_{in}$ are calculated as
$\sigma_{el}$/ $\sigma_T$ and $\sigma_{in}$/ $\sigma_T$. If P$_{el}$ $\geq$ 
R$_{4}$, an elastic collision has 
taken place.

Energy loss that occurs during elastic collision, due to target recoil, is 
calculated as 
\begin{equation}
\bigtriangleup E=\frac{m^2v^2}{m+M}-\frac{m^2vV_1\cos\delta}{m+M},
\end{equation}                      
$$
V_1=v\left[\frac{m\cos\delta}{m+M}+\frac{[M^2+m^2(\cos\delta-1)]^{1/2}}
{m+M}\right].
$$
  
Here $\delta$ is the scattering angle in the laboratory frame, $v$ and
$m$ are, respectively, the velocity and mass of the electron, and $M$
is the mass of the target particle. Differential elastic cross sections
(discussed in section 2.1) are used to obtain the scattering
angle $\delta$. Differential cross sections are fed numerically in
the Monte Carlo model at 28 unequally spaced energy points (3.2, 4.2,	5, 6,
7.9, 10, 15.4, 20, 30, 50, 100, 200, 300, 400, 500, 700, 800, 900, 1000 eV; and
1, 2, 3, 4, 5, 6, 7, 8, 9, 10 keV) and at 19 scattering angles
($0^\circ$, $10^\circ$, $20^\circ$, $30^\circ$, $40^\circ$, $50^\circ$, 
$60^\circ$, $70^\circ$, $80^\circ$, $90^\circ$,
$100^\circ$, $110^\circ$, $120^\circ$, $130^\circ$, $140^\circ$,
$150^\circ$, $160^\circ$, $170^\circ$, and $180^\circ$). 
At intermediate energies and scattering angles, values are 
obtained through linear interpolation. The energy 
$\bigtriangleup E$ is subtracted from the
energy of the primary particle. After the collision, the deflection 
angle relative to the direction ($\theta,\phi$) is obtained as 

$$
\cos\theta^{''}=\cos\theta\cos\theta^{'}-\sin\theta\sin\theta^{'}
\cos\phi^{'} ,                    
$$
\begin{equation}
\cos\phi^{''}=(\cos\theta\cos\phi\sin\theta^{'}
            \sin\phi^{'}-\sin\phi\sin\theta^{'}\sin\phi^{'}
           +\sin\theta\cos\phi\cos\theta^{'})/\sin\theta^{''},
\end{equation}
$$
\sin\phi^{''}=(\cos\theta\cos\phi\sin\theta^{'}\cos\phi^{'}
            -\cos\phi\sin\theta^{'}\sin\phi^{'}                                 
            +\sin\theta\sin\phi\cos\theta^{'})/\sin\theta^{''}.
$$

Here $\theta^{'}$, $\phi^{'}$ are the scattering angles.

If P$_{el}$ $<$ R$_{4}$ and P$_{in}$ $\geq$ R$_{4}$, an inelastic 
collision has taken place.
In this case decision has to be made on the type of inelastic 
event. 
If it is an ionization event, then a secondary electron will be generated. 
The energy of secondary electron is then calculated
with a random number R using the relation \citep{Bhardwaj09}
\begin{equation}
  T=\frac{\Gamma_S\ E_v}{E_v+\Gamma_B}[\tan(RK_1+(R-1)K_2)]+T_S
     -\left[\frac{T_A}{E_v+T_B}\right],          \label{eqn7}
\end{equation}
where
$$
K_1 = \tan^{-1}\left\{\left[\frac{(E_v-I)}{2}-T_S
    +\frac{T_A}{(E_v+T_B)}\right]
    \bigg/\frac{\Gamma_S\ E_v}{(E_v+\Gamma_B)}\right\},
$$
$$
K_2 = \tan^{-1}\left\{\left[T_S
     -\frac{T_A}{(E_v+T_B)}\right]
     \bigg/\frac{\Gamma_S\ E_v}{(E_v+\Gamma_B)}\right\}.
$$
Here $E_v$ is the incident electron energy, $\Gamma_S$, $\Gamma_A$, $T_A$, 
$T_B$, and $T_S$ are
the fitting parameters which are given in Table \ref{tab-sec-ele}, and $I$ is 
the ionization threshold. Secondary electron energy, 
calculated using equation (\ref{eqn7}), is compared with the lower cut off 
energy, which is taken as 1 eV in the current model.
If the energy of secondary electron is found to be greater than the cut off energy, then they 
are also followed in the same manner as that of a primary electron. Similarly the tertiary, quaternary, etc electrons  
are also followed in the Monte Carlo simulation.

Every single electron is followed in similar manner and number of collision
events occurred are recorded in appropriate energy bins.
The energy bin size is taken as 1 eV for the entire energy range. After each 
event, the amount of energy loss
due to the event is subtracted from the electron 
energy. After subtraction, if the electron energy is higher than cut off energy, it is again followed in the 
simulation. Electron degradation
continues until all energies are below 1 eV. The sample size is taken as 
10${^6}$ for each simulation.

\begin{table} 
\centering
\caption{Parameters for Secondary Electron Energy}
\begin{tabular}{ccccccc}
\hline
Process & I (eV) &  $\Gamma_S$ & $\Gamma_B$ & $T_S$ & $T_A$ & $T_B$ \\
\hline
CH${_4}{^+}$ & 12.99 & 30 & 10 & -0.989 & 10 & 44.52\\
CH${_3}{^+}$ & 14.24 & 23 & 28 & 0.947 & 100 & 64.52\\
CH${_2}{^+}$ & 15.2  & 7.52 & -38.1 & 2.6 & 21.8 & 44\\
CH$^+$       & 22.6  & 11   & -0.5 & -0.8473 & 0 & 0  \\
C${^+}$ & 27 & 11 & -0.5 & -1.8473 & 200 & 200 \\
H${_2}{^+}$ & 23.53 & 2 & 0.0 & -5.8473 & 0 & 0  \\
\hline
\label{tab-sec-ele}
\end{tabular}
\end{table}
\section{Yield Spectrum}

\begin{figure}
\noindent\includegraphics[width=0.75\linewidth]{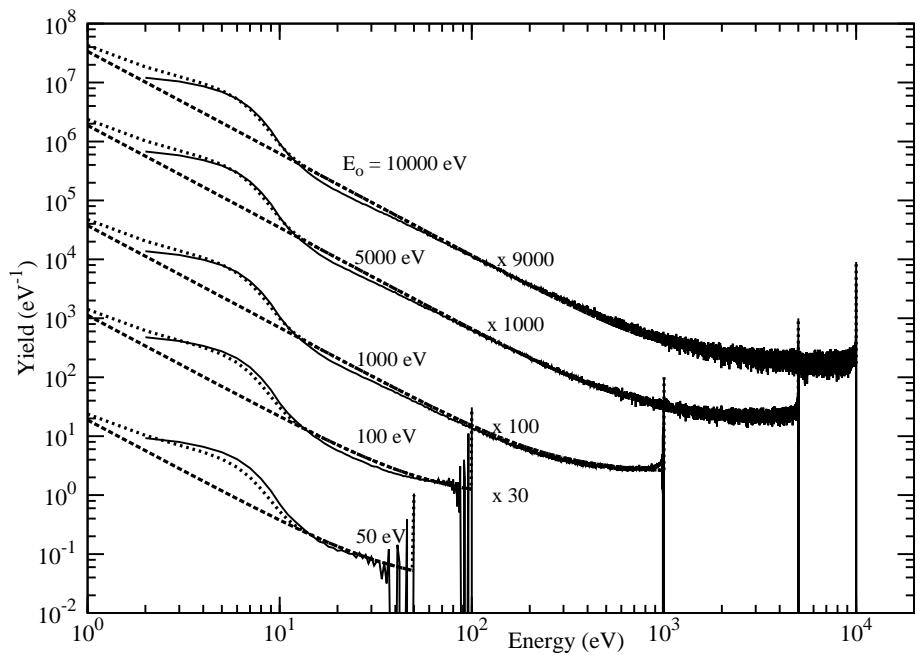}
\caption{Yield spectra for different incident energies. Solid curve shows 
numerical
yield spectra obtained using the model. Analytical Yield Spectrum (AYS), 
calculated
using equation (\ref{eqn11}), is represented by dashed curves. Dotted curve 
shows improved AYS, 
obtained by summing equations (\ref{eqn11}) and (\ref{eqn12}). Yield Spectrum 
for 10000, 5000, 1000, 100 and 50 eV are shown after multiplying with scaling factors 9000, 1000, 
100 and 30 and 1, respectively}
\label{nys&ays}
\end{figure}

A two dimensional yield spectrum, U(E,E${_\circ}$), which is a function of 
spectral energy E and incident electron energy E${_\circ}$, is obtained as the output of the Monte Carlo 
simulation. Yield spectrum gives information
about the number of energy loss events that have occurred in an energy bin and 
is defined as \citep{Bhardwaj99a, Bhardwaj09}

\begin{equation}
      U(E,E_0)=\frac{N(E)}{\bigtriangleup E}, \label{eqn8}
\end{equation}
where $N(E)$ is the number of inelastic collision events for which the
spectral energy of the electron is between $E$ and $E+\bigtriangleup E$, with $\bigtriangleup E$ 
being the energy bin width, which is 1 eV in the current
model. Figure \ref{nys&ays} shows yield spectrum at five incident energies.
If vibrational cross section measurements of \citet{Tawara92} are used in the 
energy range 1-10 eV, there is no significant change in yield spectrum. The maximum 
deviation is about to be 4.5\%.

For practical applications, yield spectrum is usually represented in the
form \begin{equation}
      U(E,E_0)=U_a(E,E_0)\ H(E_0-E-E_m)+\delta(E_0-E).  \label{eqn10}
\end{equation}
Here $H$ is the Heavyside function, with $E_m$ being the minimum 
threshold of the processes considered, and $\delta(E_0-E)$ is the 
Dirac delta function which accounts for the collision  at source energy  
E$_0$. In many atmospheric applications, it is convenient to
represent the yield spectrum, $U_a(E,E_0)$, in an analytical form 
\citep{Green77},
\begin{equation}
      U_a(E,E_0)=A_1\xi _0^s+A_2(\xi _0^{1-t}/\epsilon^{3/2 +r}) \label{eqn11}
\end{equation}
where $\xi=E_0/1000$ and $\epsilon=E/I$ ($I$ is the lowest ionization 
threshold which is equal to 12.99 eV), $A_1=0.024,\ 
A_2=4.40,\ t=0,\ r=-0.27,$ and $s=-0.085$ are 
the best fit parameters. 

The yield spectra at energy region very close to E${_\circ}$ shows rapid oscillation
 which is caused by  Lewis effect. This irregular nature
occurs due to the fact that the process of energy degradation is not 
continuous, but discrete in nature. There are only certain energies near E${_\circ}$,
which an electron can acquire. For a process with threshold E${_m}$, the
electron will suffer a minimum energy loss of 
E${_\circ}$ - E${_m}$ and this will bring down the electron energy to specific 
discrete values.
No energy value in between E${_\circ}$ - E${_m}$ can be 
acquired by the electron. To account for this effect, Heavyside 
function 
is included in the right hand side of equation (\ref{eqn10}).

The analytical yield spectrum (AYS), given by equation (\ref{eqn11}), 
well represent the numerical yield spectrum at energy values greater than the
ionization threshold (13 eV). 
To improve AYS at lower 
energies, we have used the additional function introduced by \citet{Bhardwaj09},
\begin{equation}\
        U_b(E,E_0)=\frac{E_0A_0e^{-A_5x}/A_3}{(1+e^{A_6x})^2} \label{eqn12}. 
\end{equation}
Here  $x=(E-A_4)/A_3$, and $A_0$, $A_3$,$A_4$, $A_5$ and $A_6$ are the fitting 
parameters. The values are $A_0 = 0.9$, $A_3 = 8.5 $,$A_4 = 7.0$, $A_5 = 0.001$ 
and $A_6 = 4.5$.
We have introduced the parameters $A_5$ and $A_6$ to get a better fit in 
the lower energy ($<$10 eV) region. The 
final AYS is the sum of 
equations (\ref{eqn11}) and (\ref{eqn12}). The numerical yield spectrum obtained from the model
as well as analytical yield spectrum are shown in Figure \ref{nys&ays}.

The analytical property of AYS is very useful 
in determining various property of the gas, like mean energy per ion pair and 
efficiency. The population or yield of any state j, which is the number of inelastic events
of type j caused by an electron while degrading it's energy from E${_\circ}$ 
to cut off, can be calculated using AYS as 
\begin{equation}
     J_j(E_0)=\int_{W_{th}}^{E_0} U(E,E_0)\: P_j(E)\, dE \label{eqn13}.
\end{equation}
Here ${W_{th}}$ is the 
threshold for the j$^{th}$ event and 
$P_j(E)$ is it's probability at the energy E, which is calculated as 
$P_j(E)=\sigma_j(E)/\sigma_{in}(E)$; $\sigma_{in}(E)$ being
the total inelastic collision cross section at energy E. Yield of any process 
calculated using equation (\ref{eqn13})
can be used to obtain mean energy per ion pair and efficiency.

\section{Mean Energy Per Ion Pair}

\begin{figure}
\noindent\includegraphics[width=0.60\linewidth]{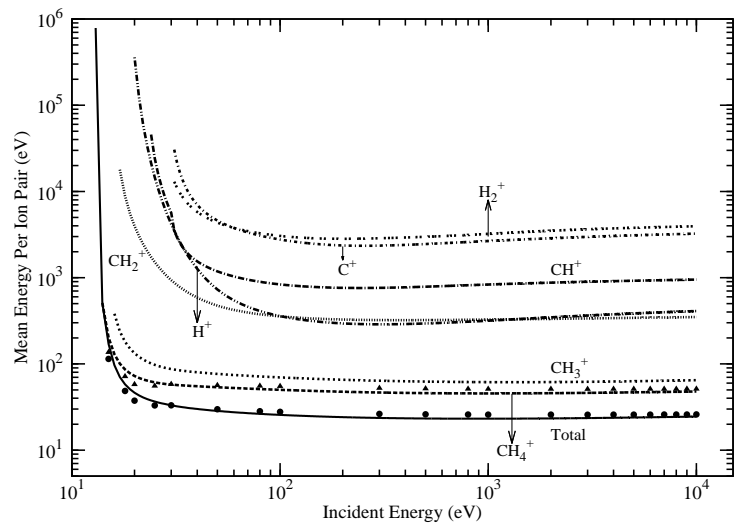}
\caption{Mean energy per ion pair for ions CH${_4}{^+}$, CH${_3}{^+}$, 
CH${_2}{^+}$,
CH${^+}$, C${^+}$, H${_2}{^+}$, H${^+}$, and neutral CH$_4$ (shown as total). 
Symbols shows 
the values calculated using numerical yield spectrum for CH${_4}{^+}$ and 
neutral CH$_4$}
\label{mepip}
\end{figure}

Mean energy per ion pair is defined as the average energy spent by an electron 
 to produce an electron-ion pair after its energy is completely dissipated. 
Its reciprocal
 gives the efficiency with which a particle can ionize the gas, and is a typical 
feature of the target species considered. 
 It is calculated as 
 \begin{equation}
     \mu_j(E_0)=E_0/J_j(E_0),    \label{eqn14}
\end{equation}
where $J_j(E_0)$ is the population of the $j^{th}$ process at the incident electron 
energy E${_\circ}$. 
At high incident electron energies, $\mu$ approaches a constant value. Figure \ref{mepip} shows
$\mu$ value calculated for neutral CH$_4$ and for the various ionization 
channels of methane.
At ionization threshold,  $\mu$ shows a very high value. As incident electron 
energy increases, population of ionization process increases as a result of which $\mu$ falls off rapidly. From about 
100 eV onwards, the curve falls off very 
slowly and attains a constant value at high incident energies.
The value of $\mu$ for CH${_4}{^+}$, CH${_3}{^+}$, CH${_2}{^+}$,
CH${^+}$, C${^+}$, H${_2}{^+}$ and  H${^+}$ ions are, respectively, 51.2 (54.9) 
eV, 68.3 (75.8) eV, 357.09 (369.9) eV, 
972.7 (855.7) eV, 3.2 (2.7) keV, 4.08 (3.1) keV and 413.06 (358.16) eV, at an 
incident energy of 10000 (100) eV. 
The mean energy per ion pair for neutral CH${_4}$ is 26 eV at 10 keV and 27.8 
eV at 
100 eV. Experimentally determined value for mean energy per ion pair, as given 
in \citet{ICRU1993}, is 27.3$\pm$0.3 eV for
incident energies $\geq$ 10 keV. \citet{Fox2008} reported a value of 31 eV, while 
\citet{Simon2011} computed a value of 
28.0$\pm$1.2 eV at an incident electron energy of 2 keV. Our value of mean 
energy per ion pair is consistent with those reported in 
previous studies. 
\section{Secondary Electron production}

\begin{figure}
\centering
\noindent\includegraphics[width=0.60\linewidth]{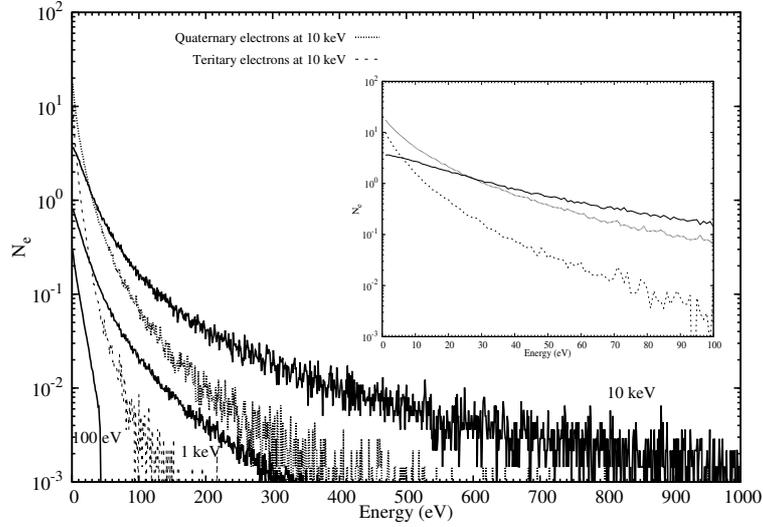}
\caption{Energy Distribution of secondary electrons for incident energies 100 eV, 
1 keV and 10 keV.
Y axis shows the number of secondary electrons produced per incident 
primary electron. Dotted and dashed
curve shows distribution of tertiary and quaternary electrons, respectively, 
for an incident energy of 1000 eV. The inset shows the energy distribution of secondary, teritary and 
quaternary electrons for an incident energy of 10 keV by zooming in the lower energy range.}
\label{sec-ele}
\end{figure}

The secondary electrons that are produced during ionization events can have a 
maximum
energy of (E-I)/2, where I is the ionization threshold. The energy of these 
electrons are
calculated using equation (\ref{eqn7}). If the energy of the secondary 
electron is greater 
than that of the cut off energy, then it is also followed in the same manner as 
that of the
primary. Similarly, tertiary, quaternary, etc electrons are also followed in
the Monte Carlo simulation. The energy 
distribution of 
secondary electrons is shown in Figure \ref{sec-ele} at few incident energies. Distribution of tertiary and 
quaternary electrons are also
shown for incident energy of 10 keV. Figure  \ref{sec-ele} shows that, each incident electron 
of energy 10 keV, at some point of its energy degradation process, produces at least one 
secondary, or teritary or quaternary electrons whose energy is $<$32 eV, which is sufficient enough
to cause further inelastic collisions.
\section{Efficiency}
During the degradation process, the electron energy is distributed among various 
inelastic processes. Efficiency of a process gives information on  what fraction 
of the incident energy is used for a particular process 
after the electron has completely degraded its energy. 
The 
efficiency, $\eta_j(E_0)$, of the $j^{th}$ process at the incident energy $E_0$ 
can be obtained as
\begin{equation}
      \eta_j(E_0)=\frac{W_{th}}{E_0}\; J_j(E_0) \label{eqn15} 
\end{equation} where $W_{th}$ is the threshold for the $j^{th}$ process. The efficiency of various processes is calculated using 
numerical yield spectrum 
as well as AYS. 

\begin{figure}
\centering
\noindent\includegraphics[width=0.60\linewidth]{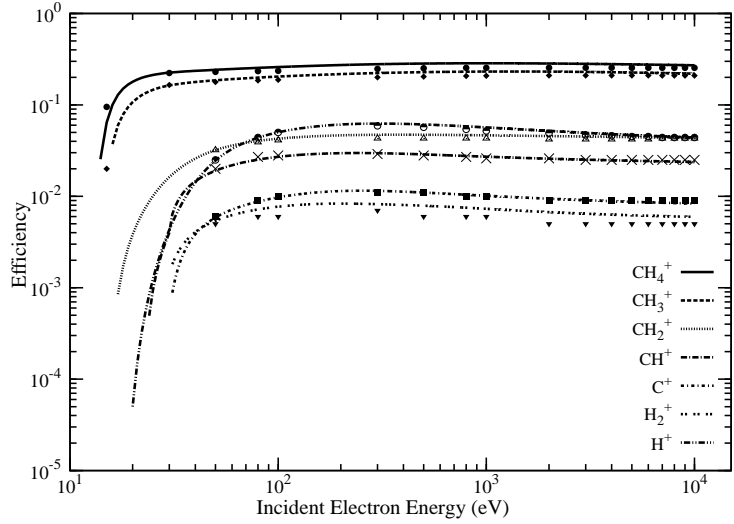}
\caption{Efficiencies of various ionization processes. Symbols
represent the efficiencies that are calculated using numerical yield spectra 
and solid
lines are efficiencies calculated using AYS.}
\label{ion-efficiency}
\end{figure}

Efficiencies of various ionization processes are shown in Figure 
\ref{ion-efficiency}. Because of its higher cross section, 
CH${_4}{^+}$ production channel has the 
highest efficiency throughout the energy range, with an efficiency of 25.3\% (23.6\%) for an incident 
electron energy of 10 keV (100 eV). The other ionization
channels  CH${_3}{^+}$, CH${_2}{^+}$, CH${^+}$, C${^+}$, H${_2}{^+}$ and 
H${^+}$ have efficiencies of
20.8\% (18.8\%), 4.3\% (4.1\%), 2.5\% (2.8\%), 0.9\% (1\%), 0.5\% (0.6\%) and 4.4\% (5\%), respectively. At 
electron energies $\geq$100 eV, there
is no significant variation in these efficiencies. But at lower energies, 
especially near the threshold region, 
ionization efficiencies fall off very rapidly. At 18 eV, 
the efficiencies for the production of CH${_4}{^+}$, CH${_3}{^+}$ and CH${_2}{^+}$ 
are 18.2\%, 9.1\% and 0.2\%, respectively. 

\begin{figure}
\centering
\noindent\includegraphics[width=0.60\linewidth]{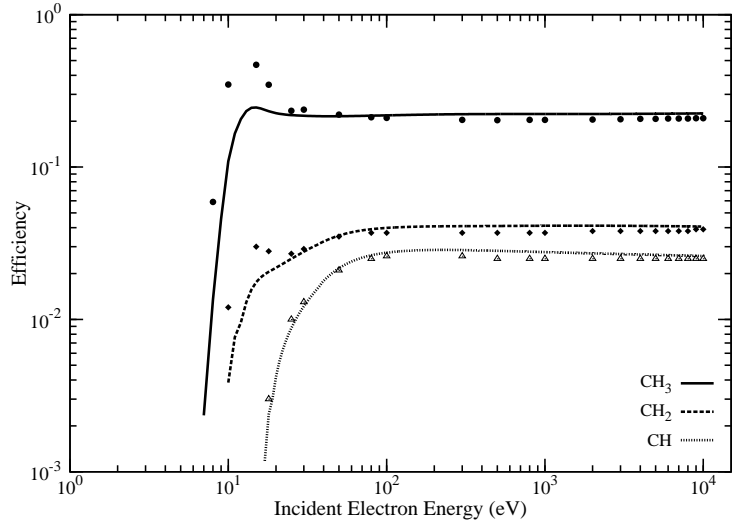}
\caption{Efficiencies of various dissociation channels. Symbols
represent the efficiencies that are calculated using numerical yield spectra 
and solid
lines are efficiencies calculated using AYS.}
\label{diss-efficiency}
\end{figure}

Figure \ref{diss-efficiency} show efficiencies of various dissociation 
channels. Since the production of CH$_3$ radical has the 
highest cross section (cf. Figure \ref{xs_channels}), 
it has the highest efficiency with a value of 20.8\% (21\%) at 10 keV (100 eV). Efficiencies of 
CH$_2$ and CH production are
3.9\% (3.7\%) and 2.5\%(2.6\%) at 10 keV (100 eV). The dissociation efficiency 
is almost constant at energies $>$100 eV. At electron energy of 30 eV, 
CH$_3$, CH$_2$ and CH dissociation channels are having efficiencies 23.8\%, 2.9\%  and 1.3\%, respectively. 

\begin{figure}
\centering
\noindent\includegraphics[width=0.60\linewidth]{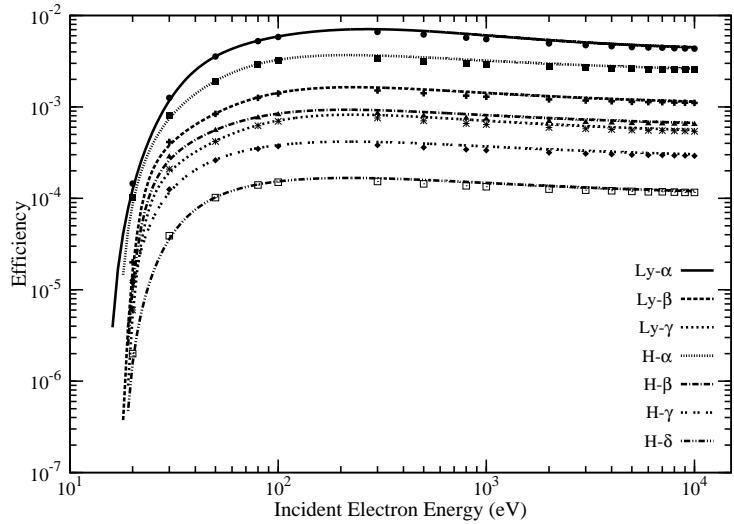}
\caption{Efficiencies of various emission channels. Symbols
represent the efficiencies that are calculated using numerical yield spectra 
and solid
lines are efficiencies calculated using AYS.}
\label{ly-ba-eff}
\end{figure}

\begin{figure}
\centering
\noindent\includegraphics[width=0.60\linewidth]{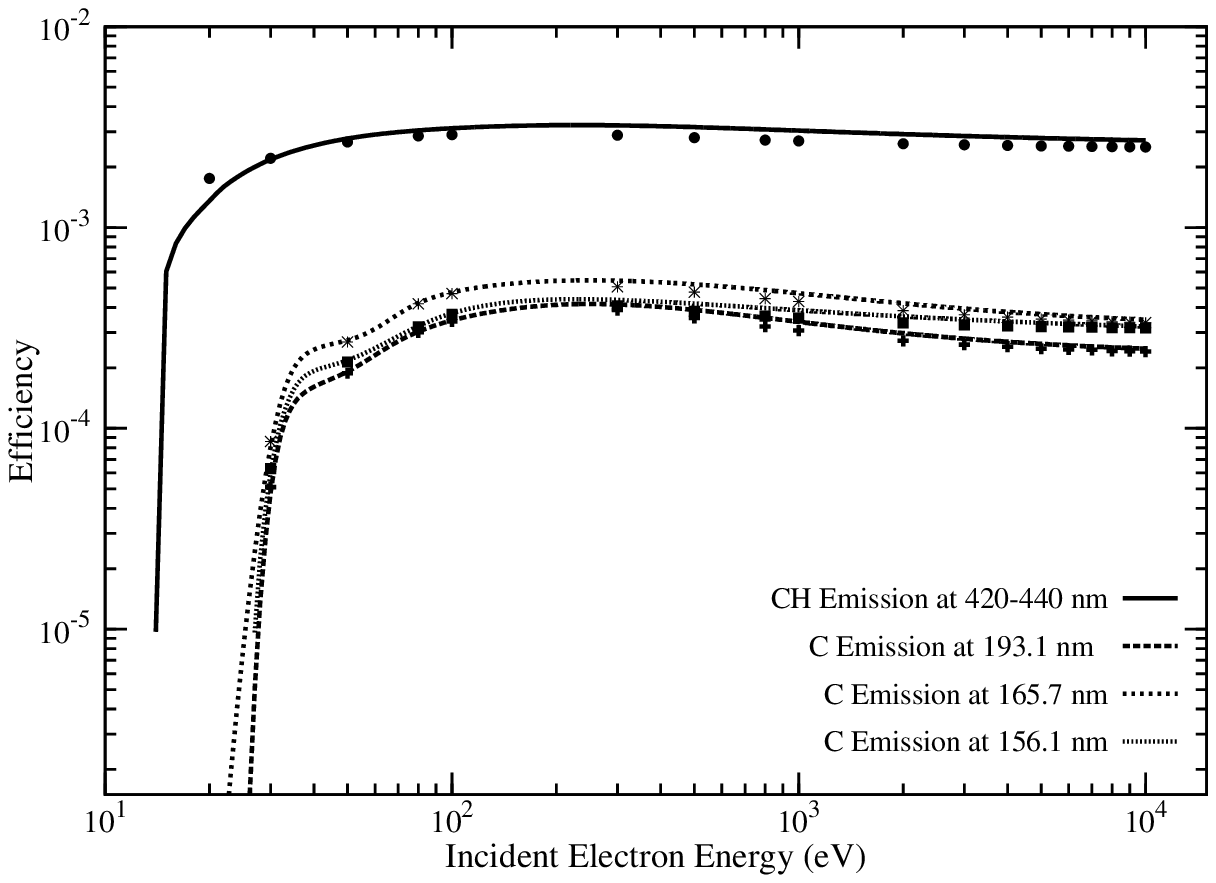}
\caption{Efficiencies of CH and various C emissions. Symbols
represent the efficiencies that are calculated using numerical yield spectra 
and solid
lines are efficiencies calculated using AYS.}
\label{ch-c-eff}
\end{figure}

Efficiencies of various emission processes are shown in Figures 
\ref{ly-ba-eff} and \ref{ch-c-eff}.
Only a small fraction of incident electron energy goes to various 
emission channels with
H Lyman-$\alpha$ emission having the highest efficiency of 0.43\% (0.58\%) at 10 
keV (100 eV). For H Lyman-$\beta$ and Lyman-$\gamma$ 
emissions, efficiencies are 0.11\% (0.14\%) and 0.05\% (0.06\%) at 10 keV (100 
eV). The CH band emission has an efficiency of
0.25\% (0.28\%) at 10 keV (100 eV). Among the various line emissions of atomic 
carbon, the 165.7 and 156.1 nm emission have almost the same efficiencies at all energies, with a 
value of 0.033\% (0.04\%)  and 0.031\% (0.03\%) at 10 keV (100 eV), respectively.
The carbon 193.1 nm emission has an efficiency of 0.02\% at 10 keV. As there is a large 
uncertainty in the value of C-line emission cross sections ($\pm$50\%),
the calculated value of efficiencies would also be uncertain by similar amount.

\begin{figure}
\centering
\noindent\includegraphics[width=0.60\linewidth]{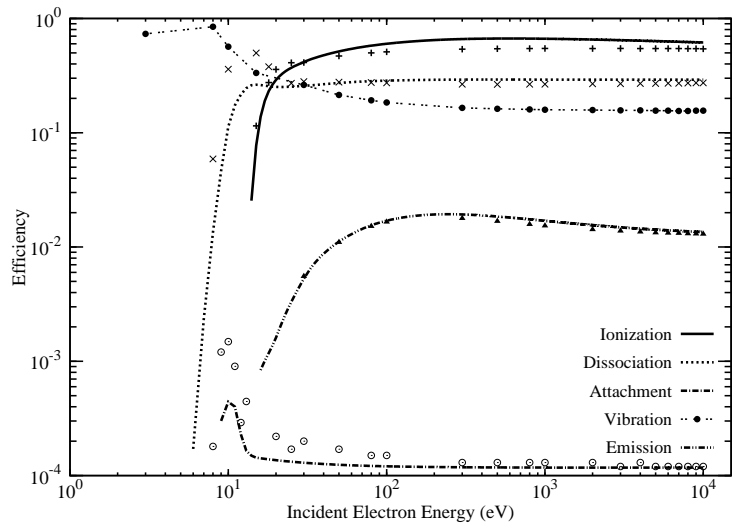}
\caption{Efficiencies of various loss channels. Symbols
represent the efficiencies that are calculated using numerical yield spectra 
and solid lines are efficiencies calculated using AYS. Vibrational efficiencies are calculated 
using numerical yield spectrum only.}
\label{all-eff}
\end{figure}

In Figure \ref{all-eff} an overall picture of efficiencies of various 
inelastic
loss processes is presented. Efficiency values calculated using both the numerical 
yield spectrum
as well as the AYS are shown; a good match is observed between the two efficiency values at energies 
greater than 
10 eV. Efficiency calculated using the AYS for energy $<$10 eV would be quite approximate 
as the AYS is not able to represent well the numerical yield spectrum in this region. Hence, the efficiency of 
vibration process shown in Figure \ref{all-eff} is calculated only using the numerical yield spectrum. Among the 
different loss processes
ionization is found to be the dominant process above 30 eV. Above 100 eV, the ionization efficiency attains a constant 
value of ~54\%. The dissociation efficiency is constant at energies above 30 
eV with a value of 27\%. 
The emission efficiency is 1.2\% (1.6\%) at 10 keV (100 eV).
All through the energy range, only a very small fraction of the incident 
electron energy is channeled into 
attachment process. The attachment efficiency peaks at ~10 eV and has a value of 
0.14\%.  

\section{Summary}
We have developed a Monte Carlo model for studying 
the degradation of 1-10,000 eV electrons in methane gas.
Analytically fitted cross sections are used as input to the model. 
The numerical yield spectra (NYS) obtained as the output of the Monte Carlo model 
includes non-spatial information about the degradation process. 
The NYS is analytically fitted using equations given by \citet{Green77}
and \citet{Bhardwaj09}. We have introduced two new parameters
to better fit the low energy ($<$10 eV) region of the NYS as described in equation \ref{eqn12}, thus 
obtaining the Analytical Yield Spectra (AYS). The AYS is used to calculate 
various parameters, like mean energy per ion pair and efficiency of various loss 
channels.
The mean energy per ion pair for CH${_4}$  has a value 26 (27.8) eV at 10 (0.1) keV. The 
energy distribution of secondary electrons for a few incident energies is presented in
 Figure \ref{sec-ele}.

Efficiency of a loss channel gives information on the amount of incident electron energy 
going into that loss process. Efficiencies are calculated 
using the AYS as well as the NYS, and
are found to be in good agreement for energy $>$10 eV. 
At energies $<$10 eV, vibration is the 
dominant loss process with an efficiency ~80\% at 8 eV. In this 
energy region, electron attachment process has an efficiency of 0.14\%,
which falls down to very small value for energy $>$20 eV. From 25 eV onwards,
dissociation process has an efficiency of 27\%. At energies higher than 100 eV, ionization
is the dominant loss process consuming more than 50\% of the incident electron energy. 
In addition to the major inelastic processes, efficiencies are 
calculated for various emissions. The H Lyman-$\alpha$ emission has the highest
efficiency among various emission channels.\\The results presented in the paper will be 
useful for modeling of aeronomical processes in the planetary atmospheres where methane
is a significant constituent. Using AYS, photoelectron fluxes in the atmosphere can be 
calculated which can be employed later on for calculating electron impact excitation or 
emission rates \citep{Jain2011, Bhardwaj99b}. Energy deposition rate can be calculated 
as a product of ionization rate and mean energy per ion pair \citep{Fox2008}.
These efficiencies can be applied to planetary atmospheres 
 for calculating volume production rates by multiplying with electron production rate and integrating over energy

\section*{Acknowledgments}
Vrinda Mukundan was supported by ISRO Research Fellowship
during the period of this work.

\section*{References}

\end{document}